\documentclass[journal]{IEEEtran}
\ifCLASSINFOpdf
\else
\fi
\ifCLASSOPTIONcompsoc
  \usepackage[caption=false,font=normalsize,labelfont=sf,textfont=sf]{subfig}
\else
  \usepackage[caption=false,font=footnotesize]{subfig}
\fi
\usepackage{url}

\usepackage{cite}
\usepackage{amsmath}
\usepackage{amsfonts}
\usepackage{amssymb}
\usepackage{bm}
\usepackage{graphicx}
\usepackage{color}

\usepackage{fancyhdr}

\usepackage{fancyhdr}
\begin{document}
%
\title{Decision-Making in Driver-Automation Shared Control: A Review and Perspectives}
%
%

\author{Wenshuo Wang, \IEEEmembership{Member, IEEE},
	Xiaoxiang Na, %
	Dongpu Cao, %
	Jianwei Gong, \IEEEmembership{Member, IEEE}, \\%
	Junqiang Xi, 
	Yang Xing,
	Fei-Yue Wang, \IEEEmembership{Fellow, IEEE}
	\thanks{Corresponding Author: Junqiang Xi.}
	\thanks{W. Wang is with the California PATH, University of California, Berkeley (UCB), CA, 94720 USA. (e-mail: wwsbit@gmail.com).}
	\thanks{X. Na is with the Department of Engineering, University of Cambridge, Trumpington Street Cambridge CB2 1PZ, U.K. (e-mail: xnhn2@eng.cam.ac.uk).}
	\thanks{D. Cao is with Mechanical and Mechatronics Engineering, University of Waterloo, 200 University Ave. West Waterloo, ON, N2L3G1, Canada. (e-mail:dongpu.cao@uwaterloo.ca).}
	\thanks{J. Gong and J. Xi are with the Department of Mechanical Engineering, Beijing Institute of Technology, Beijing, China 100081. (e-mail:gongjianwei@bit.edu.cn; xijunqiang@bit.edu.cn)}
	\thanks{Y. Xing is with the School of Mechanical and Aerospace
		Engineering, Nanyang Technological University, Singapore 639798. (e-mail:xing.yang@ntu.edu)}
	\thanks{F.-Y. Wang is  is with the State Key Laboratory for Management and Control of Complex Systems, Institute of Automation, Chinese Academy of Sciences, Beijing 100190, China, and also with the Research Center for Military Computational Experiments and Parallel Systems Technology, National University of Defense Technology, Changsha 410073, China. (e-mail:feiyue.wang@ia.ac.cn)}
}

\maketitle

\begin{abstract} 
Shared control schemes allow a human driver to work with an automated driving agent in driver-vehicle systems while retaining the driver's abilities to control. The human driver, as an essential agent in the driver-vehicle shared control systems, should be precisely modeled regarding their cognitive processes, control strategies, and decision-making processes. The interactive strategy design between drivers and automated driving agents brings an excellent challenge for human-centric driver assistance systems due to the inherent characteristics of humans. Many open-ended questions arise, such as what proper role of human drivers should act in a shared control scheme?  How to make an intelligent decision capable of balancing the benefits of agents in shared control systems? Due to the advent of these attentions and questions, it is desirable to present a survey on the decision-making between human drivers and highly automated vehicles, to understand their architectures, human driver modeling, and interaction strategies under the driver-vehicle shared schemes. Finally, we give a further discussion on the key future challenges and opportunities. They are likely to shape new potential research directions.  
\end{abstract}


%
\IEEEpeerreviewmaketitle

\section{Introduction}
\IEEEPARstart{I}{ncreasingly} powerful technologies have facilitated autonomous vehicles to prevent traffic accidents, improve traffic efficiency, and make cars available for everyone, but many social and technical obstacles remain on the road to fully autonomous driving\cite{erlien2016shared,maurer2016autonomous}. Overcoming these challenges to enable autonomous cars to drive in highly complex driving situations safely may require some time\cite{casner2016challenges}. Analogous to the levels of automation which range from complete human control to complete computer control \cite{parasuraman2000model,sheridan1978human}, the Society of Automotive Engineer (SAE) International defines five levels to describe autonomous vehicles and have been adopted by the U.S. National Highway Traffic Safety Administration (NHTSA) in Washington, D.C., as the government's template. As a transition to autonomous vehicles, \textit{partially automated} car, in which the human driver and automated driving agent\footnote[1]{An automated driving agent refers to as a well-designed automatic controller in automated vehicle systems.} \textit{share} and complete a driving task, becomes a compromise plan before the era of fully autonomous vehicles \cite{sae2014taxonomy}. Level 3 of automated driving, called conditionally autonomous driving, enables vehicles to mutually transit driving modes between fully automated driving and full manual control\cite{russell2016motor}, but this would degrade the vehicle performance, primarily when transferring the vehicle control authority from car to a driver \cite{gordon2015automated,lv2017analysis}. Because it requires a transition period for the driver to be resumed in the driving process\cite{eriksson2017takeover}, which can often pose difficulties when the driver has not been actively engaged in the driving process to reacquire situation awareness\cite{russell2016motor}. Research demonstrates that vehicle automation\footnote[2]{Automation is a technology that actively selects data, transforms information, makes decisions, or controls processes\cite{lee2004trust}.} harms mental workload and situation awareness \cite{endsley1999level,stanton2005driver} and that reaction times increase as the level of automation increases\cite{young2007back}. Another kind of automated driving is called semi-automated driving, in which the automation system does not take full authority from the driver, that is, the driver should keep eyes firmly on the road, though the driver feet off the pedal, hand off the steering wheel\cite{gordon2015automated}. However, humans will be bored and distracted during a low-level supervision task \cite{llaneras2013human,gordon2015automated} and will show over-trust \cite{inagaki2010theoretical,lee1992trust}, neglect\cite{goodrich2001experiments}, and complacency\cite{billings1996human} on automated driving systems, which requires a long time to resume control from the automation system in critical situations\cite{eriksson2017takeover}. Therefore, Gordon and Lidberg \cite{gordon2015automated} hold that semi-autonomous driving does not alleviate the regular task of anticipating traffic hazards.

\begin{figure}[t]
	\centering
	\includegraphics[width = 0.48\textwidth]{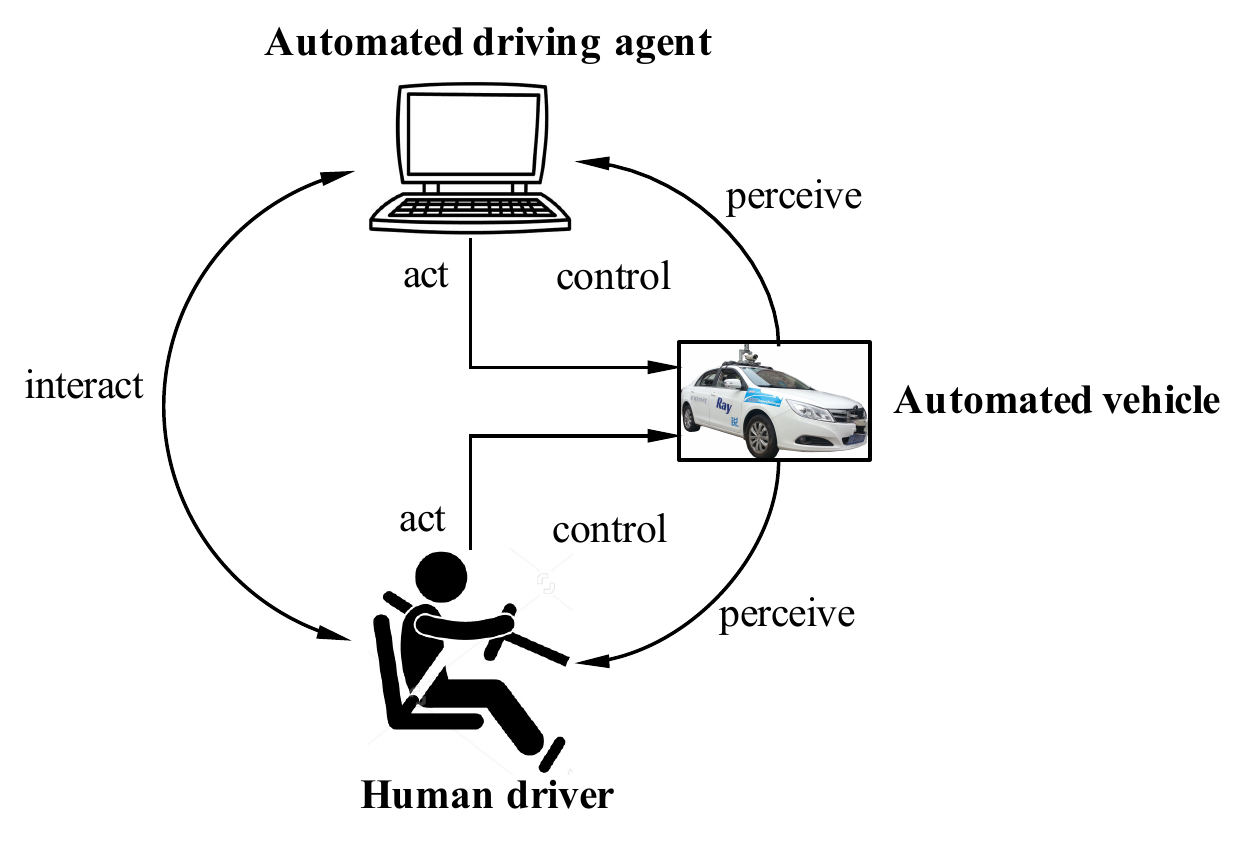}
	\caption{Illustration of the interactions between the human driver and the automated driving agent\cite{flemisch2008automation,waard2007,meyer2018gesture}.}
	\label{fig:driver_vehicle_interaction}
\end{figure}

Human interaction with automated driving agents is a kind of human-robot interaction (HRI), and one of the most influential concepts for HRI was \textit{supervisory control} \cite{sheridan1992telerobotics}. Supervisory control usually involves a human supervisor setting up the goals while the automated driving agent applying control actions to achieve the goals \cite{abbink2018topology}. However, practically it would be difficult to establish a sharing of control between the human supervisor and the automated agent in this context\cite{ferrell1967supervisory}. Sheridan and Verplank \cite{sheridan1978human} defined two sub-concepts to explain the idea of supervisory control further. One is \textit{traded control} which requires that both human and computer are active at the same time, and the other one is \textit{shared control} in which at one time the computer is active, at another the human is\cite{inagaki2003adaptive}.

The shared control scheme presents a tractable paradigm to tackle the driving authority transition issues \cite{mulder2012sharing,saito2018control} in Level 3. It features achieving a continuous authority transition between the human driver and automated driving agent. Though the authorized levels of automated driving were issued in 2016, the underlying shared-control concept has already been introduced in other fields \cite{sheridan1992telerobotics,abbink2018topology,inagaki2003adaptive} and will be the sharp end of cooperation between agents\cite{flemisch2016shared}. Each agent in the shared control scheme can take their advantages when performing a specific driving task \cite{sheridan2002humans}. It is well known that manual control is prone to human errors. On the other hand, fully automated tasks are currently subject to wide-ranging limitations in decision-making and situation-awareness. To exploit full potentials of both of human and automation while overcoming the barriers of car-to-driver transition, Mulder, \textit{et al}.\cite{mulder2012sharing} presented an entirely different control scheme -- \textit{shared control systems}\footnote[3]{The definition of shared control is slightly different over different research fields because there is no single definition for shared control that is used
across application domains. More detailed descriptions are referred to see Section II in review paper \cite{abbink2018topology}.}. The human driver and the automated driving agent continuously share and cooperatively complete a specific driving task, thereby allowing drivers to enjoy driving while keeping in control consistently. Moreover, the shared-control scheme can synergize innate human capacities and technological capabilities to enable us to
realize our full potential\cite{ren2016rethinking}. Previous research experimentally demonstrates that keeping driver's haptic control authority in the loop with continuous haptic feedback to the driver not only outperforms the conventional binary switches between supervisory and manual control \cite{petermeijer2015should} but also reduces distraction on a secondary task \cite{griffiths2005sharing} and drivers' workload\cite{tada2017simultaneous}. Fig. \ref{fig:driver_vehicle_interaction} presents a driver-vehicle system where the human driver and automated driving agents cooperatively share and achieve the same driving task. The shared control over manual control has shown the advantages in many applications such as lane keeping assistance \cite{blaschke2009driver,saito2016driver} and steering assistance system\cite{wang2017human,wang2015human,ercan2017predictive,tada2017simultaneous}. 

To some extent, driver-vehicle shared control is a kind of driver assistance system, and from this point of view, which includes three categories: \textit{perception enhancement} (e.g., informational assistance), \textit{action suggestion} (e.g., decision or action selection), and \textit{function delegation} (e.g., action implementation) \cite{de2011preparing,parasuraman2000model,payre2016fully,hoc2009cooperation}. 
The first two types have been reviewed in \cite{casner2016challenges}, except for the third type, in which both human drivers and automated driving agents can exert their inputs to vehicles simultaneously to carry out a specific task such as the active steering assistance systems. A well-designed driver-vehicle shared control system should allow all engaged agents to know each other very well \cite{fisher2016humans}, which requires addressing the following fundamental research questions:
\begin{enumerate}
	\item What kind of role should the human driver act in the shared control system with changing situations?
	\item How to allocate the driving responsibility and authority according to the ability of two agents?
	\item How to on-line evaluate the respective trust levels among drivers and automated driving agents?
	\item What are the temporal scales of human adaptation and learning in changing situations?
	\item What novel system identification techniques for driver state and intent exist that could allow us to study time-varying and possibly nonlinear shared control systems?
\end{enumerate}

Although many works of literature have been done for specific topics, there is no paper to review and discuss these research questions comprehensively. To bridge the gap, we provide an overview of the field of decision-making scheme design and human driver modeling in shared control systems, by reviewing more than 200 closely related literature covering the keywords: shared control, driver model, shared cognitive control, self-driving, and human-automation/robot/computer interaction. Instead of reviewing rigorous mathematical algorithms of decision-making and controller design, we mainly focus on the scheme design of decision-making, human driver modeling, and the open issues with potential solutions in shared control systems, thus benefiting researchers working on the considered topic. Some other partially related literature on psychology and ergonomics is only involved and referred without in-depth discussion because of page limitation. Section II describes the underlying architectures of driver-vehicle shared control systems. Section III reviews the decision-making of two agents in the driver-vehicle shared control system from the state-of-the-art literature. Section IV details human driver modeling. Section V shows and discusses some open-ended, challenging, inevitable scientific questions. Section VI gives further discussion and conclusion.

\section{Shared Control Architectures}

\begin{figure}[t]
	\centering
	\subfloat[]{\includegraphics[width = 0.48\textwidth]{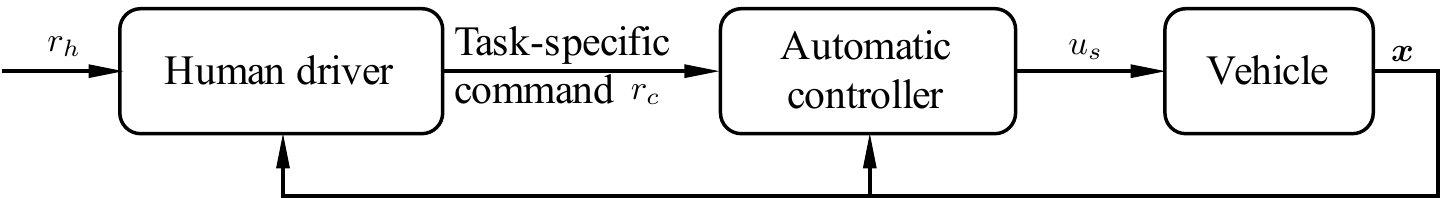}}\\
	\subfloat[]{\includegraphics[width = 0.48\textwidth]{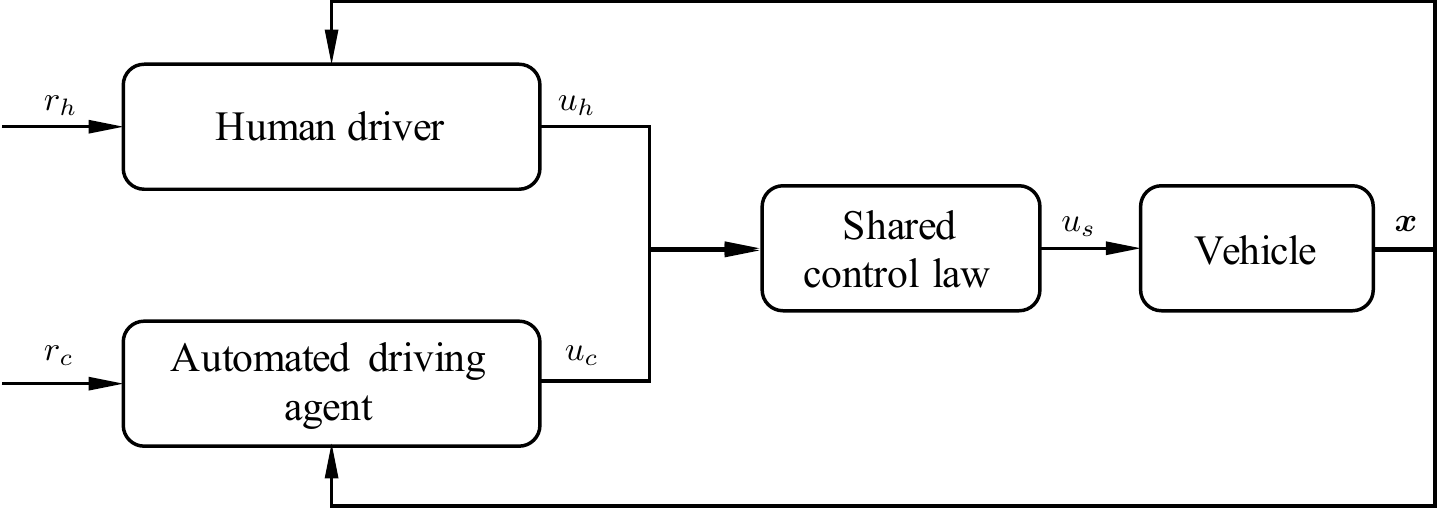}}\\
	\caption{Two kinds of shared control architectures: (a) Task-level shared control and (b) Servo-level shared control. $r_{h}$ and $r_{c}$ represent the expected trajectories to be followed for human driver and automated driving agent, respectively; $u_s$ is the input of vehicle, $u_h$ and $u_c$ are the human driver's operation output and automated driving agent's output, respectively.}
	\label{fig:shared_control_structure}
\end{figure}

Before giving an insight review, we first discuss the architecture of the driver-vehicle shared systems. A shared-control system consisting of a human driver and an automated driving agent refers to as a two-agent system that is capable of accepting and executing commands from a human driver, or an automated driving agent, or a combination of the two\cite{venkataraman1993shared}. According to the role that the human driver plays in the driver-vehicle shared systems, the shared control can occur at two different levels: the task level and the servo level (as shown in Fig. \ref{fig:shared_control_structure}).

\subsection{Task-Level Shared Control}
In the task-level shared control\footnote[4]{The task-level shared control refers to as a shared-control system that allows human drivers to decompose a whole driving task into subtasks and allocate some of them to the automated driving agent (namely, share at the task level\cite{venkataraman1993shared} rather than the servo level), which is slightly different from the traded control \cite{abbink2018topology,sheridan1992telerobotics,sheridan1978human}.} scheme\cite{rieger1982allocation,venkataraman1993shared,tan2019guidance}, human drivers usually act as a guide and deliver a task-specific command to the automated driving agent. Namely, the human driver can allocate subtasks to the automated driving agent while authorizing other subtasks to achieve a complete driving task. After being informed the subtasks, the automated driving agent will perform the subtasks based on the current situation condition and predefined algorithms to cooperatively achieve the whole driving task together with the human driver. The task-level shared control scheme fully exploits the strength in both machines and humans, which could alleviate the driving burden of a human. A very intuitive example is that human drivers manually activate the adaptive cruise control (ACC) systems when driving on the highway\cite{marsden2001towards} to allocate longitudinal control authority to the ACC agent while authorizing lateral control by himself/herself.

\subsection{Servo-Level Shared Control}
Servo-level control in the driver-vehicle system usually focuses on servo control. The control input ($ u_{s} $) to a vehicle is typically the combination of human drivers' operations ($ u_{h} $) and automated driving agents' output ($ u_{c} $), as shown in Fig. \ref{fig:shared_control_structure}(b). In the servo-level shared control, differing from the task-level shared control situations where the automated driving agent will \textit{take over} the task-specific control, human drivers will always be engaged in the control process of vehicle movement at the servo level. The combination of outputs from human drivers and automated driving agents should be well designed. Analogous to human-robot shared control, researchers\cite{borroni2018weighting,li2018shared,sentouh2018driver,guo2018predictive,guo2018hazard} combine them intuitively  by

\begin{equation}\label{eq:proportion}
u_{s} = \lambda u_{h} + (1-\lambda)u_{c}
\end{equation}
where $ \lambda \in [0,1]$ is the weighted coefficient to adjust the proportion of $ u_{h} $ and $ u_{c} $ in $ u_{s} $. The allocation of driving authority for a human driver and an automated driving agent is determined by $ \lambda $ which can be either fixed or continuously adaptive. Pure human control is achieved when $ \lambda = 1 $ and pure automatic control when $ \lambda = 0 $. Therefore, the value of $ \lambda $ can impact the shared control scheme by determining if the driver is presented in the loop. This parameter can be modulated manually (e.g., in \cite{saito2016shared}) and automatically based on the driver's states in certain situations in which the driver needs assistance to perform more rapidly and safely (e.g., in \cite{li2018shared}). The design of the $ \lambda $ will be discussed in Section III.

\begin{figure}[t]
	\centering
	\includegraphics[width = 0.45\textwidth]{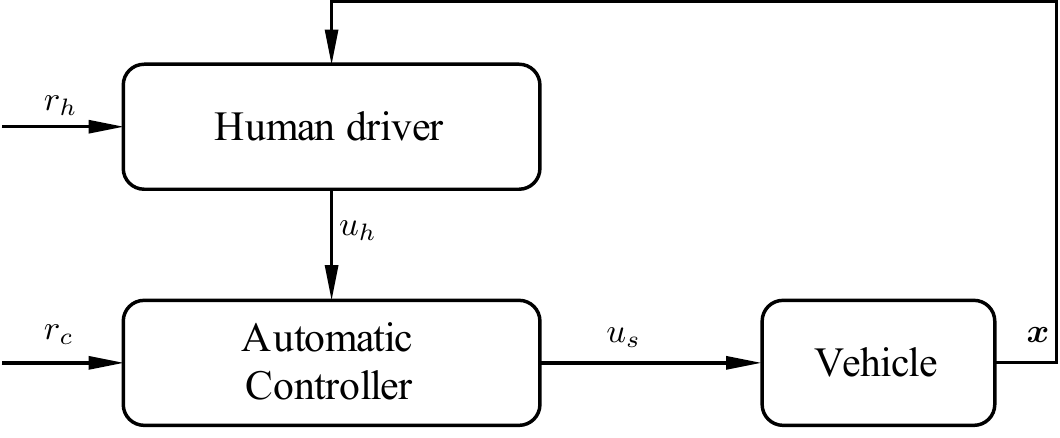}
	\caption{Illustration of an indirect shared control scheme\cite{li2017indirect}, also called ``input-mixing shared control''\cite{abbink2010neuromuscular}.}
	\label{fig:SC_structure2_1}
\end{figure}

\begin{table*}[t]
	\centering
	\caption{Driver-in-the-loop (DIL) Model Using State-Space Representation Towards the Applications in Vehicle Dynamic Control}
	\begin{tabular}{ccccc}
		\hline
		\hline
		Year & Reference & Application & Vehicle model & Driver model\\
		\hline
		2001 &\cite{chen2001differential} & Rollover prevention & 3 DOF yaw-roll model & UMTRI driver model \cite{macadam1989mathematical} \\
		2007 & \cite{chen2007design} & Steering system control& 12 DOF vehicle model &  Single-point preview driver model \cite{macadam1981application}\\
		2008 & \cite{pick2008mathematical} & -- & Linear yaw/sideslip vehicle model & Path-following controller with NMS \\
		2009 & \cite{sentouh2009sensorimotor} & Path-following task & 2 DOF bicycle model & 3-level driving steering control model\\
		2010 & \cite{sentouh2010toward} & Shared lateral control & 2 DOF bicycle model &  Two-point visual preview model \\
		2013 &\cite{saleh2013shared} & DIL stability analysis & 2 DOF vehicle model & Two-point visual driver model with NMS\\
		2014 & \cite{soualmi2014automation} & LKA system& Nonlinear road-vehicle model & Preview driver model\\
		2014 &\cite{khosravani2014development} & Path-following task & 2 DOF bicycle model & Preview model with time delay\\
		2014 & \cite{mashadi2014path} & Path-following task& 8 DOF nonlinear vehicle model & Modified preview model\\
		2016 &\cite{nash2016development} & Steering system control & - & Driver model incorporated sensory dynamics \\
		2016 & \cite{merah2016new} & LKS and LDA & 3 DOF vehicle model & Fuzzy controller \\
		2016 & \cite{koh2016integrated} & DIL simulation & Carsim & Speed-steering control driver model\\
		2016, 2017 &\cite{schnelle2016personalizable,schnelle2017driver} & Collision avoidance & 10 DOF vehicle model & Compensatory and anticipatory model\\
		2017& \cite{ercan2017predictive} & Steering assistance system & Nonlinear vehicle model & Preview model with NMS \\
		2014,2017 & \cite{nguyen2017driver,oufroukh2014integrated} & LKA system &2 DOF bicycle model & Two-point visual control model\cite{salvucci2004two}\\
		{2019} & \cite{huang2019datadriven} & {Path-following task} & {2 DOF bicycle model} & {Two-point visual preview model} \\
		\hline
		\hline
		\multicolumn{5}{l}{DOF -- Degree of freedom; NMS -- Neuromuscular systems; LKA -- Lane keeping assistance; LDA -- Lane departure assistance;} \\
		\multicolumn{5}{l}{Carsim -- Vehicle simulation software.}
	\end{tabular}
	\label{Table:1}
\end{table*}

Different from linearly combining the human driver's input with the automated agent's input by (\ref{eq:proportion}), some probabilistic models \cite{simpson1999automatic} have also been proposed by explicitly taking into account the uncertainty in the interaction and modeling this combination as a joint probability distribution\cite{ezeh2017probabilistic}.

The approaches of combining $ u_h $ and $ u_c $ can differ from each other in different servo systems. As a consequence, the servo-level shared control can be further divided into \textit{direct} shared control and \textit{indirect} shared control\cite{li2017indirect,tian2020indirect}. 

\subsubsection{Direct shared control} {The direct shared control allows both human drivers and automated driving agents to simultaneously exert actions on a control interface}, of which the output remains the direct input to vehicle systems, as shown in Fig. \ref{fig:shared_control_structure}(b). Such systems are usually \textit{haptic shared control} \cite{abbink2010neuromuscular} since both human drivers and automated driving agents will directly influence the inputs on {the} haptic control interface (e.g., steering handwheel and brake/throttle pedals). {Also, human drivers can even percept} the assistance torque applied by automated driving agents through the shared haptic interface. A general architecture of haptic shared control can be seen in \cite{boehm2016architectures}.

\subsubsection{Indirect shared control}
Differing from the direct shared-control scheme, the indirect shared control scheme shapes the input to the controlled vehicle system by mixing the out of control interface (usually as a result of human contributions) and output of the automated driving agent \cite{abbink2010neuromuscular,wang2017human,wang2015human}, formulated by $ u_s = g(r_h, r_c, u_h)$ as shown in Fig. \ref{fig:SC_structure2_1}. A typical application in the human-vehicle systems is steer-by-wire (SBW) systems\cite{huang2019shared} which estimates the driver's desired steering angle from the driver operations and then generates and applies the steering angle directly to front wheels.

\section{Decision-Making in The Servo-Level Shared Control Scheme}

This paper mainly focuses on the decision-making in servo-level shared control scheme rather than the task-level shared control scheme. A well-described model of the entire driver-vehicle system can offer one a better understanding of how each subsystem works and what the relationship between them is. Some researchers hold that driver modeling should be integrated into the road-vehicle system for control purposes to improve the mutual understanding between the driver and automation \cite{saleh2013shared,soualmi2014automation}. Hence, a driver-in-the-loop (DHL)\footnote[5]{The ``loop'' can refer to an information processing control loop (i.e., attentive to driving task) or a sensory-motor control loop (i.e., vehicle control), or both\cite{louw2015engaging}. Here, the ``loop'' refers to as a sensory-motor control loop.} vehicle model \cite{saleh2013shared,nguyen2017driver,wang2017gain} is usually incorporated into a shared control system and then describes the complete DIL systems from a control perspective. 

For this purpose, first, the road-vehicle dynamic model {is usually formulated using} the state-space representation,

\begin{equation}\label{eq:road_vehicle_model}
\dot{\boldsymbol{x}} = A\boldsymbol{x} + B(u_{c} + u_{h}) + B_{w}w
\end{equation}
where $ \boldsymbol{x} $ is the road-vehicle state, $ A $ is the road-vehicle system matrix, $ B $ is the input matrix from road-vehicle system, $ B_w $ is the system disturbance matrix, $ u_c $ is the controller input, $ u_h $ is the human operation input, $ w $ is a system disturbance. Then the human driver is formulated from a control perspective as 

\begin{equation}\label{eq:drivermodel}
u_{h} = \mathcal{H}(r_h, \boldsymbol{x})
\end{equation}
where $ r_h $ is the driver's desired trajectory or reference trajectory. $ \mathcal{H}(r_h,\boldsymbol{x}) $ represents human driver model and can be formulated from control perspective \cite{plochl2007driver}, stochastic perspective \cite{zeng2017stochastic,wang2017development,wang2017evaluation}, or cognitive perspective \cite{salvucci2006modeling}, which will be discussed in Section IV. The driver model towards the applications to steering system control (or lateral control, path-following control) is the preview driver model and its extensions (see review article \cite{plochl2007driver}) since it is easy to integrate them into the state-space-based vehicle model. Human drivers output their operations and apply to the vehicle by comparing the desired trajectory and the current trajectory through their internal model\cite{macadam2003understanding,cole2017occupant}. The internal model can estimate the current/upcoming states of the subject vehicle  and trajectories of surrounding objects (e.g., vehicles, pedestrians, and bicycle users).  Substituting (\ref{eq:drivermodel}) into (\ref{eq:road_vehicle_model}), the driver-vehicle model (i.e., DIL model) can be formulated by a new state-space representation

\begin{equation}\label{eq:incorp_DIL}
\dot{\boldsymbol{z}} = A_{DIL}\boldsymbol{z} + B_{h} u_{h} + B_{c}u_{c} + B_{w}w
\end{equation}
where $ \boldsymbol{z} $ is the augmented state consisting of human driver model state and road-vehicle system state, $ A_{DIL} $ is the driver-vehicle system matrix, $ B_{h} $ is the human input matrix, and $ B_{c} $ is the controller input matrix.
The common DIL models based on the state-space representation can be found in \cite{chen2007design,nguyen2017driver,chen2001differential,mashadi2014path,nash2016development,saleh2013shared,li2015continuous} and some popular cases are also listed in Table \ref{Table:1}. {The state-space DIL model provides an analytical way to assess the stability of the shared control system\cite{saleh2013shared,huang2019datadriven} and an standard way to design controllers\cite{bencloucif2019cooperative,sentouh2018driver}. A well-designed decision-maker and controller for the shared control system should assist human drivers driving safely and smoothly while without causing any conflicts with human drivers.} Based on the incorporated DIL model (\ref{eq:incorp_DIL}), given a desired trajectory/reference and the human driver operations, {the optimal controller input can be obtained by solving an optimal problem in the general form
\begin{equation}
u_{c}^{*} = \arg \min_{u_c} J(\cdot)
\end{equation}
where $ J(\cdot) $ is the objective function that could encompass human driver's input and other constraints (e.g., vehicle dynamics and human driver's physical limitations).} To solve the optimization problem between a human driver and an automatic controller, one of the biggest challenges is to formulate the relationship between $ u_c $, $ u_h $, and $ u_s $, e.g., allocation of the control authority, $ \lambda $. In what follows, the shared control strategies can be formulated according to prior knowledge or dynamic programming. Thus, two ways to design the decision-making strategies are listed: rule-based and game theory-based.
\subsection{Rule-Based}
For the rule-based method, one direct way is to use (\ref{eq:proportion}) with the requirement of designing $ \lambda (t) $. Most research predefined a different kind of rule according to prior knowledge, which can be roughly grouped into three categories and discussed as follows.

\subsubsection{Piecewise Function}{Due to the complexity of dynamic environment and disturbance, an intuitive way to design $ \lambda(t) $ is using rule-based \textit{piecewise function}. The piecewise function is primarily developed to tackle the shared-control issues in robotics such as wheelchair and industrial robots \cite{jiang2016shared,li2017human} and then is introduced to intelligent vehicles afterward because of its robustness and practicality in terms of controller design\cite{jiang2017robust,jiang2018shared,jiang2017shared}.} For instance, Jiang and Astolfi \cite{jiang2017shared,jiang2016shared} defined three space sets to divide the reachable set into three parts by judging the level of safety ---  safe, close, and dangerous. Correspondingly,  a three-level piecewise function consisting of the safe $ \mathcal{R}_{s} $, hysteresis $ \mathcal{R}_{h} $, and dangerous subsets $ \mathcal{R}_{d} $ was proposed to design $ \lambda (t)$:

\begin{equation}
\lambda(t) = 
\begin{cases}
	f_{1} (\boldsymbol{x}(t), u_{h}(t),u_{c}(t)), \ \mathrm{if} \ (\boldsymbol{x}(t), u_{h}(t)) \in \mathcal{R}_{s} \\
	f_{2}(\boldsymbol{x}(t), u_{h}(t),u_{h}(t)), \ \mathrm{if} \ (\boldsymbol{x}(t), u_{h}(t)) \in \mathcal{R}_{h}\\
	f_{3}(\boldsymbol{x}(t), u_{h}(t),u_{c}(t)), \  \mathrm{if} \ (\boldsymbol{x}(t), u_{h}(t)) \in \mathcal{R}_{d} \\
\end{cases}
\end{equation}
If human behaves ``dangerously'' then the feedback controller (i.e., automated agent) is active and resumes the control authority; if human behaves ``safely'' then the vehicle only responses to human's operations; if human behaves in ``hysteresis'' set, then the vehicle runs under a predesigned shared-control law.

Besides, the piecewise function is an analytical way to take account human factors and driving situations in the light of its practical integration with human's prior knowledge. For example, Li, \textit{et al}. \cite{li2018shared} utilized two piecewise functions to assess driving situations and vehicle performance separately based on their empirical knowledge and then fused these piecewise functions as the $ \lambda(t) $. Saito, \textit{et al}. \cite{saito2018control} used an exponential function of the steer-wheel torque applied by the human driver to estimate how much torque assistance the assisting agent should deliver.

\subsubsection{Exponential Function}The second approach to obtain a seamless and smooth $ \lambda $ is to {use the family of the \textit{exponential function}:

\begin{equation}\label{equation_7}
\lambda(t) = \frac{1}{1-e^{\alpha_{1}(\frac{\alpha_{0}}{\alpha_{\mathrm{safe}}}-\zeta(t))}}
\end{equation}
where $ \zeta(t) $ is the activity factor, $ \alpha_{0} $ and $ \alpha_{1} $ are the tuning parameters, and $ \alpha_{\text{safe}} $ is the parameter to guarantee safe and model convergence. The exponential function has been widely used in human-robot shared control, and then introduced to tackle control authority transfer and allocation in human-vehicle shared control afterward. For example, Sentouh \textit{et al}.\cite{sentouh2018driver} integrated the discrete driver state (distraction) into the exponential function to obtain a continuous shared control factor $ \lambda(t) $. Wang, \textit{et al}. \cite{wang2017human} designed a shared steering control law using {an} exponential function with considering different driving styles to improve vehicle performance and reduce drivers' workload when taking curve negotiation. However, {a religious setting} should be made when applying this exponential function, since the derivative of Equation (\ref{equation_7}) could be discontinuous when the denominator switches between positive and negative. 

In order to determine a continuous weighted coefficient ($ \lambda $) for the shared control systems, researchers also combine the \textit{piecewise function} and \textit{exponential function} to evaluate the safety level for decision-making in human-robot shared control\cite{li2017adaptive}.} {This kind of shared control strategy has been carefully borrowed to improve human-vehicle shared control performance. For example, Sentouh \textit{et al}. \cite{sentouh2018driver} utilized the function in the format of (\ref{equation_7}) to obtain a continuous authority allocation factor from a discrete driver drowsiness monitoring factor. Besides, some probabilistic shared control strategies towards complex, dynamic environments were also proposed by modeling both the human's intentions and the automated agent as a probabilistic function to improve the shared control performance with a exponential function\cite{bencloucif2019cooperative,tran2018human}.}

\begin{figure}[t]
	\centering
	\includegraphics[width = 0.33\textwidth]{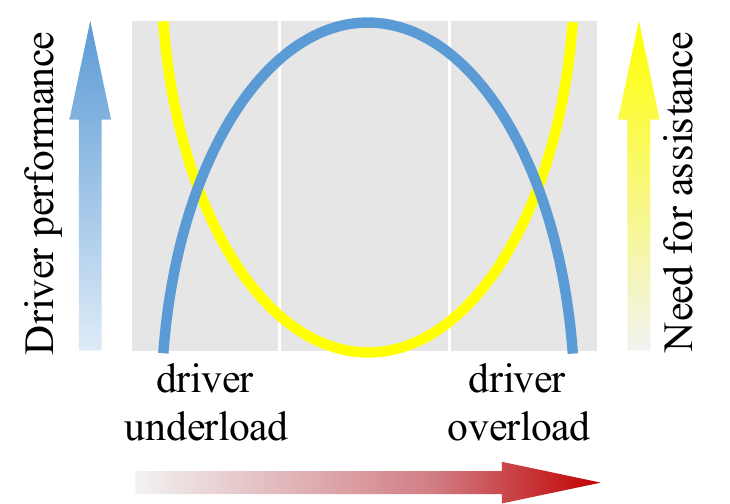}
	\caption{U-shape representation of driver's need for assistance\cite{flemisch2010towards,gordon2015automated,hoeger2008highly}.}
	\label{fig:U_shape}
\end{figure}
\subsubsection{U-Shape Function}
Another approach is to directly compute how much assistance the driver needs based on the \textit{U-shape function}, i.e., the relationship between drivers' workload and performance as well as the need for assistance, as shown in Fig. \ref{fig:U_shape}.  For instance, Nguyen, \textit{et al}. \cite{nguyen2017driver,nguyen2017sensor} designed {an} assistance torque $ T_{a} = \mu(a) T_{s}$ to reduce drivers' workload and improve vehicle performance using the U-shape function of drivers' activity $ \mu(a) $, where $ T_s $ is the required input torque from the vehicle, and $ a $ is the drivers' activity denoted as steering angle. Oufroukh and Mammar \cite{oufroukh2014integrated} also proposed a similar computation model to compute the assistance torque using U-shape representations during lane keeping or obstacle avoidance maneuver.

\subsection{Game-Theory-Based}

\begin{figure}[t]
	\centering
	\includegraphics[width = 0.48\textwidth]{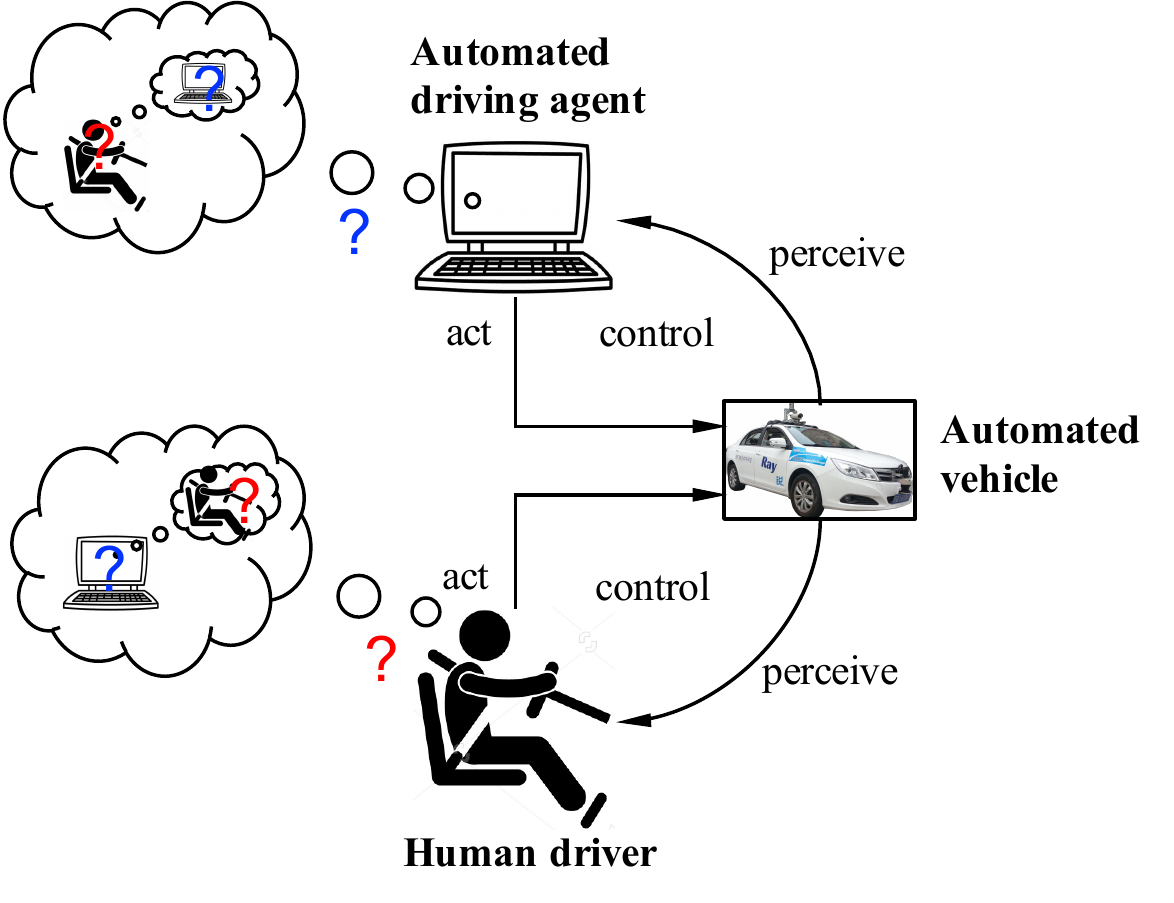}
	\caption{Dynamic games between a human driver and the automated driving agent.}
	\label{fig:humanvehicleGT}
\end{figure}

\begin{figure}[t]
	\centering
	\includegraphics[width = 0.48\textwidth]{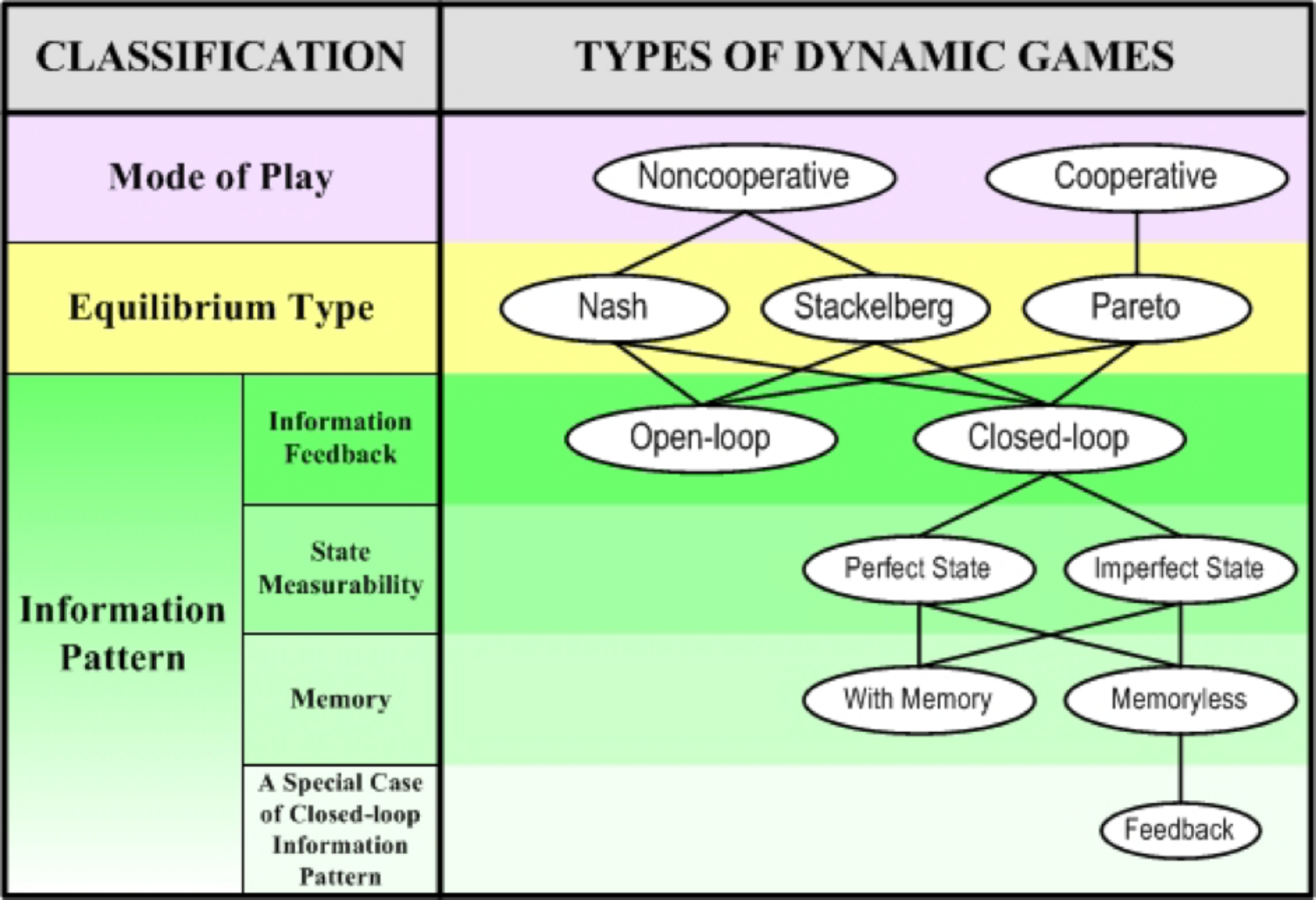}
	\caption{Classification of dynamic games\cite{na2015game}.}
	\label{fig:game_theory}
\end{figure}

\begin{table*}[t]
	\centering
	\caption{Game-Theoretical Applications in the Shared Control Between Human Driver and Automated Driving Agent}
	\begin{tabular}{cccc}
		\hline
		\hline
		Year&Reference & Application & Types of game theory\\
		\hline
		2011 &\cite{tamaddoni2011optimal} & Stability/yaw control & Noncooperative \\
		2014 &\cite{dextreit2014game} & Energy management & Noncooperative \\
		2013, 2015&\cite{na2015game,na2013linear} & Collision avoidance control & Noncooperative/cooperative\\
		2017 &\cite{na2017application} & Active steering control &Noncooperative \\
		2017 &\cite{flad2017Cooperative} & Driver assisted steering & Noncooperative \\
		{2019} & {\cite{li2019shared}} & {Obstacle avoidance} & {Noncooperative} \\
		{2019} & {\cite{ji2019shared}} & {Lane change} & {Noncooperative} \\
		\hline
		\hline
	\end{tabular}
	\label{table:gametheory}
\end{table*}
{Differing from the rule-based method where a predefined shared control law $ \lambda(t) $ is adopted, some researchers treated human drivers and automated driving agents in the shared-control system as two-player with dynamic interaction}, as shown in Fig. \ref{fig:humanvehicleGT}. The two-player assumption enables game-theoretic approaches to be available to tackle the relations between human drivers and automated driving agents.

Game-theoretic approaches have been widely applied to tackle the dynamical decision-making problem where two or more agents make decisions that influence one another welfare\cite{thomas2012games} such as vehicle-to-vehicle \cite{bahram2016game,wang2015game,talebpour2015modeling}, grid-to-vehicle \cite{tushar2012economics,wang2014game}, collision avoidance at an intersection \cite{hafner2013cooperative}. Applications of game theory in modeling road user behaviors and traffic or transportation can refer to the review literature \cite{elvik2014review,bazzan2014review}. Na and Cole made a comprehensive classification for dynamic games as shown in Fig. \ref{fig:game_theory}. Depending on the interactive type between human drivers and automated agents,  the dynamic games between two agents can be classified into noncooperative and cooperative games\cite{shoham2008multiagent}, as illustrated below.
\begin{itemize}
	\item \textbf{Noncooperative:} Drivers and automated driving agents consider themselves as individuals and concentrate on {pursuing their interest}. More specifically, in the noncooperative game theory, the strategy type of driver and automated driving agent can be derived using Nash equilibrium and Stackelberg equilibrium. A Nash equilibrium emerges in situations, where drivers and automated driving agents derive {their strategies} by considering others' strategies, and they act simultaneously. A Stackelberg equilibrium emerges in situations, where one agent (i.e., human driver or automated driving agent) is the leader, and the other one serves as a follower.
	\item \textbf{Cooperative:} Drivers and automated driving agents have a sense of collectivity and attempt to enter into a binding agreement of interest, where the goal of each agent is identical, and their strategies are derived from global optimality\cite{fele2017coalitional}.
\end{itemize}

Usually, the two agents are assumed to be rational with individual objectives \cite{na2015game,na2017application,flad2017Cooperative,na2019modelling}. The vehicle is controlled by a human driver and an automated driving agent, modeled as 

\begin{equation}
\dot{\boldsymbol{x}}(t) = f(t, \boldsymbol{x}(t), u_{c}(t), u_{h}(t)).
\end{equation}
The goal of a human driver and automated driving agent is to minimize their objective function:

\begin{equation}\label{eq:human_obj}
u_{h}^{*} = \arg \min_{u_h} J_{h}(t, \boldsymbol{x}(t), \boldsymbol{x}^{\mathrm{ref}}_{h}(t), u_{c}^{*}, u_{h})
\end{equation}

\begin{equation}\label{eq:control_obj}
u_{c}^{*} = \arg \min_{u_c} J_{c}(t, \boldsymbol{x}(t), \boldsymbol{x}^{\mathrm{ref}}_{c}(t), u_{c}, u_{h}^{*})
\end{equation}
where $ J_{h}(\cdot) $ and $ J_{c}(\cdot) $ are the objective functions of the human driver and automated driving agent, respectively; $ \boldsymbol{x}^{\mathrm{ref}}_{h}(t) $ and $ \boldsymbol{x}^{\mathrm{ref}}_{c}(t) $ are the desired/reference trajectories of the human driver and automated driving agent, respectively. The objective functions (\ref{eq:human_obj}) and (\ref{eq:control_obj}) both depend on the vehicle state $ \boldsymbol{x} $ and the two agent's inputs, $ u_c $ and $ u_h $. Here, (\ref{eq:human_obj}) and (\ref{eq:control_obj}) can be same or different. Table \ref{table:gametheory} lists some literature using game theory to tackle the driving authority between human drivers and automated driving agents. We notice that the means of modeling relations between a human driver and automated driving agent using game-theoretic approaches has been introduced since 2011.

Taking the case where  (\ref{eq:human_obj}) and (\ref{eq:control_obj}) are symmetric for example, it is necessary to priorly know the human driver's input $ u^{*}_{h} $ if we want to solve the optimization problem in (\ref{eq:control_obj}). However, the human driver's actions {also strictly depend} on the automated driving agent's actions, as shown in (\ref{eq:human_obj}). Different assumptions can result in different game-theoretic schemes:
\begin{itemize}
	\item \textit{Noncooperative Nash} scheme, where the automated driving agent will compensate for human drivers' erroneous actions to imitate the possible opposite effects, and \textbf{simultaneously}, human drivers will also estimate action applied by the automated driving agent, and to neutralize its influence by further changing her/his inputs;
	\item \textit{Noncooperative Stackelberg} scheme, where the automated driving agent will compensate for human drivers' erroneous actions to imitate the possible opposite affects, and \textbf{sequentially} human drivers will generate actions when they have a very well knowledge of the actions applied by automated agents, and vice versa; 
	\item \textit{Cooperative Pareto} scheme, where both human drivers and automated driving agents {try to account for} each other's desired trajectory, and \textbf{simultaneously} will react to both each other's actions.
\end{itemize}

The Nash and Stackelberg equilibriums can be employed to solve the closely coupled optimization problem in driver-vehicle interactions. Either analytical or approximated solutions can be derived as the human driver's and the automated driving agents' control actions. When the approximated solution is concerned, the following two expressions hold:

\begin{equation}\label{eq:appr1}
 u_{h}^{*}  \approx \tilde{u}_{h} 
\end{equation}
\begin{equation}\label{eq:appr2}
u_{c}^{*}  \approx \tilde{u}_{c} 
\end{equation}
For example, Flad, \textit{et al}. \cite{flad2017Cooperative} introduced an approximated Stackelberg solution\cite{chen1972stackelburg} to solve the problem between a human driver and ADAS controller (i.e., automated driving agent) by treating one of them as a leader and the other one as a follower. In \cite{li2015continuous}, Li, \textit{et al}.  designed a continuous role adaptation of human-robot shared control using the Nash-equilibrium, where human and the automatic controller (i.e., automated driving agent) can simultaneously exert control to the robot,  instead of directly using the Stackelberg solution \cite{flad2017Cooperative,na2017application}. An adaption law was also designed by comparing the difference between measured human drivers' inputs and predefined Nash equilibrium. {Besides, the game theory could also integrate the uncertainty from human drivers and external factors into the shared control system \cite{ji2019shared} with a stochastic dynamic programming solver.

Although the above mentioned two typical methods offer us ways to design the shared control strategies between a human driver and automated driving agents, the two agents could also fail to cooperate when a wrong estimation of current states regarding driving situation perception and human driver intent occurs. Also, a poor-designed shared control strategy could even bring four main adverse effects\cite{hoc2001towards,hoc1998cognitive}: loss of expertise, complacency, trust, and loss of adaptivity. Therefore, studies dealing with human-vehicle cooperation have to consider the human's characteristics such as uncertainty according to reduce the conflicts with the driver.  In what follows, we will discuss about modeling human drivers from different perspectives.}


\section{Human Driver Modeling}

\begin{figure}[t]
	\centering
	\includegraphics[width = 0.48\textwidth]{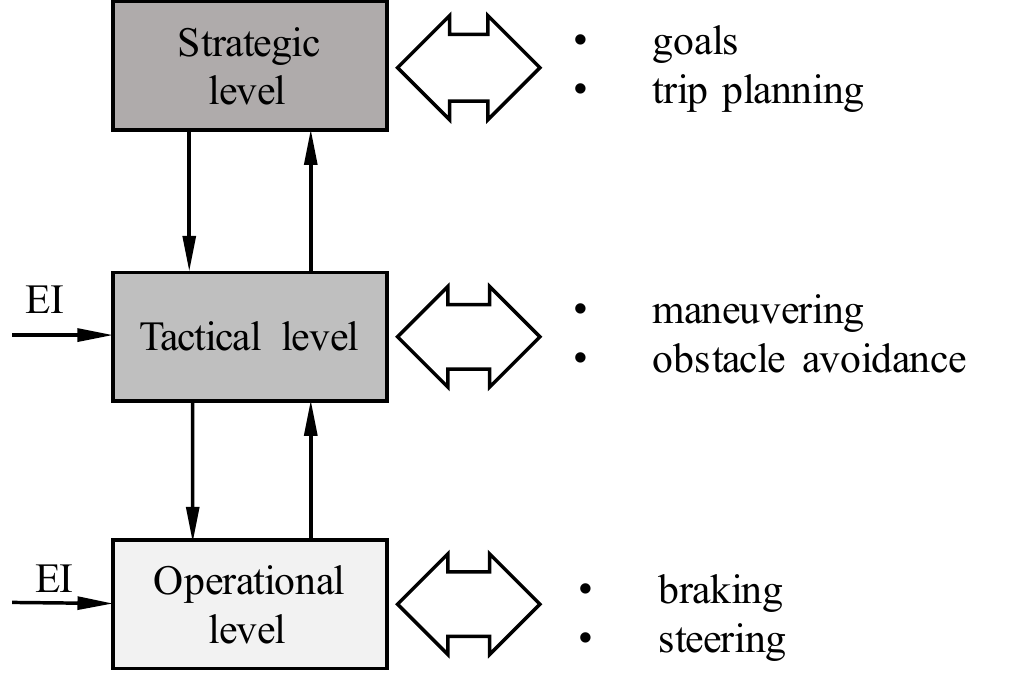}
	\caption{A hierarchical structure for modeling human drivers when driving\cite{michon1985critical}. EI = Environment inputs.}
	\label{fig:driver_model}
\end{figure}

In order to obtain well-designed shared control interactions between a human driver and a highly automated vehicle, Norman \cite{norman1990problem} stated that human must always be in control, must be actively engaged and adequately informed, and that human and automated vehicle must understand each other's intents correctly in the complex driver-vehicle systems. Therefore, understanding and modeling human drivers' sensory dynamics, cognition processes, hidden states, and operation characteristics are equally important as dynamic vehicle systems for driver-vehicle shared control systems. Human drivers usually complete a driving task at three levels \cite{michon1985critical,rasmussen1986information} (Fig. \ref{fig:driver_model}): \textit{strategic} level, \textit{tactical} level, and \textit{operational} (control) level. The first two levels involve cognition while the third level involves execution. According to {the previous review} of all literature, we introduce and discuss the driver model from its functional modules, modeling approaches, and driver's intent inference as well as state detection.

\subsection{Functional Module}
\subsubsection{Sensory Dynamics}
{A recent} review in \cite{nash2016review} demonstrates that drivers' sensory dynamic characteristics play an important role in driver-vehicle system design, making them interact more {friendly and safe}. Driver's sensory dynamics used for vehicle speed and direction control mainly include\cite{nash2016review} (ranked by the level of importance for a regular driving task\cite{macadam2003understanding}):

\begin{itemize}
	\item Visual --- The visual system is the human driver's only means of detecting {the future road trajectory}. A general review of overall literature demonstrates that visual information is the highest significant in normal driving process, {accounted for about} 90\% of all sensory information\cite{macadam2003understanding}. Driver's visual information (e.g., eye gaze\cite{doshi2009roles,salvucci2002time}) can also reflect driver's underlying intentions (e.g., lane change intents\cite{li2016bayesian}, readiness to take over from automation\cite{zeeb2015determines}), mental or physical states (e.g., fatigue detection\cite{mandal2017towards}), and upcoming actions\cite{mars2012we}.
	\item Vestibular and kinesthetic --- Human drivers usually use the motion information (e.g., acceleration and rotation of vehicles) derived from vestibular and kinesthetic channels as a supplement to visual information for control task, which could also contribute to the human combined arm-trunk motion\cite{mars2003vestibular}.
	\item Somatosensory --- The somatosensory information used in control of vehicle steering and speed mainly includes tactile and haptic information such as pressure on the gas/brake pedal \cite{abbink2011measuring} and steering torque on the steering handwheel\cite{ercan2017predictive}.
	\item Auditory --- The auditory information usually is used as a supplementary cue within a multi-channel environment in a normal driving process. 
\end{itemize}
The sensory dynamics have their physical characteristics such as time delay, perception limitations, and coherence zones because of human's physical, biochemical, mental  limitations and the human's ability to perceive and process information. A driver model which integrates human drivers' sensory dynamics could help understand humans, thus for improving comfort, safety and driveability of driver-vehicle shared control systems and reducing the incompatibility\footnote[6]{Compatibility here is referred to as the quality describing the fit or match \cite{flemisch2008automation} between a human driver and automated driving agents, regarding the outer (e.g., interfaces) and inner (e.g., cognition) interactions \cite{coll1989cognitive} between them.} or negative interference \cite{hoc2001towards} between {the} human driver and automation. The haptic information on {the} steering wheel has been widely used in {the} shared steering control systems to reduce the driver's mental/physical workload and to improve driving skills \cite{nishimura2015haptic,inoue2016enhancing,petermeijer2015should}. A sensorimotor driver model was developed to improve the shared-control performance by considering both visual and kinesthetic perception and including compensatory and anticipatory processes \cite{sentouh2009sensorimotor}.

\subsubsection{Cognition} {John A. Michon has mentioned the cognitive driver model} from a critical view of introducing the behavioral sciences and psychology to understand the human decision-making processes \cite{michon1985critical}.
{Much of the decision-making generated} by human drivers is over discrete actions, such as choosing whether/when/how to lane change\cite{nilsson2016if}. To capture these discrete features, researchers modeled human drivers' high-level tactical behavior (e.g., speed selection and decision making) and strategic behavior (e.g., route planning and navigation). One of the most utilized means is based on the ``Adaptive Control of Thought-Rational (ACT-R)\cite{anderson2013architecture}'' cognitive where the discrete nature of drivers' control actions is captured from a cognitive perspective. For example, Salvucci, \textit{et al}. developed an integrated cognitive path-following driver model \cite{salvucci2001toward} and lane-change driver model\cite{salvucci2007lane} using the combination of the ACT-R cognitive architecture and perceptual-motor process. 

Some researchers also developed cognitive driver models based {on on-hand knowledge} or with new insights in experimental data. Misener, \textit{et al}. \cite{misener2000emergence} developed a cognitive car-following model to avoid rear-end crashes with a stopped lead vehicle by fusing current knowledge derived from experimental data. Liang, \textit{et al}. \cite{liang2007real} developed a system to detect driver distraction in real-time by cognitively analyzing three indicators: {how to define distraction}, which data were input to the model, and how the input data were summarized. Liang also demonstrated that combining cognitive and visual distractions can improve vehicle performance than each of them\cite{liang2010combining}. {Much more literature} on modeling and analyzing drivers' cognitive distraction can see references \cite{li2015predicting,hirayama2016classification,harbluk2007road,strayer2013measuring}. A review paper for modeling driver behaviors in a cognitive architecture refers to \cite{salvucci2006modeling,cole2017occupant}.

Based on the cognitive driver model, a cognitive {assist} can be potentially provided in the context of driver-vehicle shared control to reduce harmful interference between two agents. Cai and Lin \cite{cai2012coordinating} proposed a coordinating cognitive {assist} to determine \textit{when} {an assist should be} provided and {\textit{how much} assistance} to be supported for a steering assist control system. The cognitive assistance was divided into three stages to assist drivers in acquiring information, analyzing information and making a decision, and implementing action\cite{parasuraman2000model}. 

\subsubsection{Neuromuscular-Skeletal Dynamics}
Existed research has demonstrated that a good understanding of neuromuscular-skeletal dynamics has constraints upon the dynamics
of perception-action coupling \cite{carson1996neuromuscular} and is {significantly vital for a well-designed} shared control system, for example, avoiding subtle conflict between {a} human driver and automated agent\cite{abbink2012importance}. For human drivers, the neuromuscular-skeletal dynamics in human-vehicle systems mainly include {the arm and foot}, representing the lateral and longitudinal control, respectively.

For the dynamic properties of a driver's holding {a} steering wheel, Pick and Cole \cite{pick2007dynamic,pick2006neuromuscular} investigated the effect of the driver opposing {a} constant offset torque and the effect of the driver co-contracting {the} muscles, and found that both actions will increase the stiffness and damping of {the} arms. A linear model of the neuromuscular system, muscles, limbs, and {the} vehicle was then set up and applied to a driver simulator\cite{hoult2008neuromuscular}, a guideline to shared control \cite{abbink2010neuromuscular}, a path-following driver–vehicle model\cite{cole2012path}, and a driver lateral control model\cite{bi2015development,pick2008mathematical}. {Besides, the characteristics of driver neuromuscular dynamics are different from each other and affected by steering systems (e.g., active and passive) and hand positions on the steering wheel \cite{lv2018characterization}, which should be considered when designing a controller. }

For the dynamic properties of a driver's low leg and foot hitting the gas or brake pedal, researchers mainly focus on the ankle-foot neuromuscular dynamics, which enables us to gain insight into both responses to visual feedback and haptic feedback. Abbink, \textit{et al}. \cite{abbink2011measuring} investigated the biomechanical properties of the ankle-foot complex (i.e., admittance) \cite{pronker2017estimating} and developed a driver support system that uses continuous haptic feedback on the gas pedal to inform drivers of the separation to the lead vehicle.

\subsection{Modeling Approaches}
\subsubsection{Control-Theoretic Driver Model}
Understanding the potential and {the limits of a tightly coupled} driver-vehicle-road system are not trivial. Many kinds of driver model towards the application in vehicle dynamics control can be found in review articles \cite{plochl2007driver,wang2014modeling,brackstone1999car}, in which literature before 2014 was listed, including lateral and longitudinal driver models. Table \ref{Table:1} lists some popular driver models toward vehicle dynamics control applications and indicates that the single/two-point visual preview driver models or their extensions are highly preferred since they are easy to be integrated with vehicle models. However, {one of the most significant limitations} is that the single/two-point visual preview driver models assume that human drivers' references or desired trajectories are exactly known, which is not always available in {the} real application. {Besides,} for such two-point visual preview models, drivers' {essential} physical characteristics such as the reaction time delay, anticipatory and compensatory can be reflected, but not cognitive characteristics such as neuromuscular dynamics of driver arms and decision-making processes. For example, the control-theoretic driver models (see review \cite{plochl2007driver}) usually ignore issues of whether or how to perceive the model inputs from the external environment and how drivers correctly interact with other automatic controllers\cite{salvucci2006modeling} via visual, haptic, or auditory sensors\cite{macadam2003understanding}. 

Most control-theoretic driver models have been used in the driver-vehicle shared control systems with a game-theoretic scheme. {With this purpose, the human driver and the automated driving agent should be first modeled, which makes it feasible to estimate each other's actions.} One of the most popular approaches for modeling automated driving agents is using control theory (e.g., model predictive control) since it can describe drivers' ability to predict vehicle's future states \cite{prokop2001modeling} based on their internal model or individual's driving skills\cite{qu2015switching}. Na and Cole \cite{na2017application,na2015game} applied the combination of the distributed MPC and linear {quadratic dynamic} optimization (LQDO) to formulate human drivers and automated driving agents. The objective function with quadratic structure is also widely used to design an automated driving agent considering human drivers' forthcoming motion primitives \cite{flad2017Cooperative} or drivers' haptic inputs on the control interface\cite{li2015continuous}. 
\subsubsection{Learning-Based Driver Model}
Though the driver models as mentioned above could describe and predict drivers' behavior, actions, and states with relatively satisfied performance, they did not concern the dynamic, stochastic decision-making processes of driver behavior, which requires models capable of connecting temporal and spatial processes.  For this purpose,  some researchers also utilized learning-based approaches to deal with highly nonlinear properties of driver behaviors, such as neural networks \cite{morton2017analysis,olabiyi2017driver}. Research in  \cite{wang2017development,wang2017learning,wang2019learning,wang2018learn} demonstrates that the Markov models combined with Gaussian mixture models achieve a satisfied performance of capturing driver intent and action. Besides, the Bayesian inferences\cite{wang2017driving,taniguchi2016sequence}, autoregressive exogenous (ARX)\cite{okuda2013modeling}, and deep learning\cite{wang2017capturing} were also developed. These learning-based approaches highly depend on the collected training data{, and} some of them are data-hungry such as deep neural networks\cite{wang2017much}.

\subsection{Intent Inference and State Detection} {Correctly inferring drivers' intents and states is profoundly essential to} design an automatic controller capable of delivering an adequate input not only to follow/track the desired trajectories { but also} to avoid intrusive interventions between the human driver and the automated driving agent\cite{rauch2009importance}. 
\subsubsection{Intent Inference}
Steering wheel, as a direct interface, allows human drivers and automated driving agents {to} act and exchange information in a simultaneous and continuous way \cite{mars2014analysis,abbink2012haptic}. Therefore, human drivers' intents can be directly captured through the torque applied on the steering wheel by drivers.  For example, in a torque-based steering assistance system\cite{ercan2017predictive}, the automated driving agent uses sensors to obtain the steering torque forced by drivers, and inversely, human drivers can also react to his/her haptic perception information from the steering wheel. Nguyen, \textit{et al}. \cite{nguyen2017driver,nguyen2017sensor} utilized the torque applied to the steering wheel by drivers as an indicator to compute how much assistance torque an automated driving agent should provide. Li, \textit{et al}. \cite{li2015continuous} proposed a continuous adaption law for the human-robot shared control system to determine the automated driving agent's role (i.e., leader or follower) by comparing the measured torque applied by the human to the predefined Nash equilibrium computed through optimal control techniques. 

In addition to using the torque applied on the steering wheel by a human, the dynamic neuromuscular analysis of driver arms or legs can also provide a guideline for shared control design of a steering system \cite{ercan2017predictive,abbink2012haptic,cole2012path,pick2008mathematical,abbink2010neuromuscular} and the gas/brake pedal control\cite{abbink2011measuring,abbink2012haptic}. For instance, in order to reduce the intrusive intervention between human drivers and automated systems, Ziya, \textit{et al}. \cite{ercan2017predictive} modeled human drivers' steering behavior by combining the neuromuscular response of drivers and the desired steering angle that was a function of vehicle states and road geometry. Moreover, the impedance of a haptic torque was used as an indicator of the drivers' intents. Some researchers also designed a guidance torque to assist drivers to keep the vehicle in the driving lane \cite{ercan2017modeling} and to improve vehicle safety for fatigue-related driver behavior \cite{wang2017effect}.

{Much literature} on modeling human drivers' neuromuscular dynamics of steering behavior has been found in \cite{cole2012path,nash2016development,pick2006neuromuscular}. The model of driver's arm and steering dynamics usually combines with a path-following control model, obtaining a linear driver model with neuromuscular dynamics \cite{pick2008mathematical}:

\begin{equation}\label{eq:neuromuscular}
(J_{\mathrm{d}} + J_{\mathrm{s}})\ddot{\theta}_{\mathrm{s}} + (B_{\mathrm{d}} + B_{\mathrm{s}})\dot{\theta}_{\mathrm{s}} + (K_{\mathrm{d}} + K_{\mathrm{s}}){\theta}_{\mathrm{s}} = T_{\mathrm{m}} - \frac{M_{\mathrm{T}}}{n_{\mathrm{s}}}
\end{equation}
where $ J_{\mathrm{*}} $ are the inertia, damping  and stiffness of $ * $, with $ * $ being driver arm or steering systems; $ n_{\mathrm{s}} $ is the steering gear ratio; $ M_{\mathrm{T}} $ is the torque arising from the lateral forces and self-aligning moment; $ T_{\mathrm{m}} $ is the muscle torque; and $ \theta_{\mathrm{s}} $ is the steering wheel angle. Model (\ref{eq:neuromuscular}) has been used to infer drivers' intents. For example, the changes of damping and stiffness of driver's arms can reflect whether the automatic controller's outputs satisfy drivers' desired trajectory or the conflict level between drivers and automated driving agents \cite{pick2007dynamic,abbink2010neuromuscular,boehm2016architectures}. Instead of using indirect cues (e.g., steering angle, vehicle dynamics), direct human-observing cues (e.g., body gesture, head, hands, feet, and gaze direction) can also be used to predict driver intent\cite{tawari2014continuous,tran2012modeling,yuen2016looking,rangesh2016long}.

{In addition to directly using the haptic information}, driver intent can also be inferred and predicted according to dynamic driving environments such as peripheral vehicles\cite{guo2015multimodal} and vehicle position to lane edges\cite{gaikwad2015lane}. The dynamic Bayesian networks (DBN) \cite{kasper2012object,sivaraman2014dynamic}, Markov decision processes (MDP) \cite{wang2017learning,tran2012modeling}, and partially observable Markov decision processes (POMDP) \cite{gindele2015learning} are one obvious place to start, assuming that one can extract the underlying latent processes of driver behavior which is a dynamic and stochastic process. {Besides, the correlation between human driver gaze behavior and steering moments will decrease while increasing the allocation ratio of control authority for intelligent driving agents \cite{wang2018relationship}, indicating that driver gaze behavior could be used to infer driver's intent and avoid conflict.}

\subsubsection{State Detection}
Correctly detecting driver state (e.g., sleepiness, drowsiness, fatigue, distraction, impairment) offers an opportunity to make {a practical} decision of authority allocation, thus improving vehicle safety. For example, Saito, \textit{et al}. \cite{saito2016driver} proposed a dual control scheme for lane-keeping assistance systems by detecting the driver's sleepiness level using eye blinking frequency and facial information. More driver intent detection and inference using direct human-observing cues can refer to the review paper\cite{ohn2016looking,mandal2017towards}.

Visual distraction or cognitive distraction have been investigated by combining vehicle state\cite{li2015predicting,hirayama2016classification,saito2016driver}, drivers' visual state \cite{liang2010combining,li2015predicting,liang2007real,harbluk2007road}, and operations\cite{li2015predicting,li2017visual,harbluk2007road}. Answers to the question of how to measure a driver's cognitive distraction have been given in \cite{strayer2013measuring}. Learning-based approaches such as deep sparse autoencoders\cite{liu2017visualization}, deep belief networks or DBNs \cite{gindele2015learning}, support vector machines (SVM) \cite{liang2007real,liu2016driver} have been widely used to detect and classify driver distraction.

\section{Future Challenges and Opportunities}
Due to the limited ability to understand human drivers, many open-ended questions still exist in driver-vehicle shared control systems. This section will present and discuss some open-ended, challenging, inevitable questions regarding the shared control strategies, the trust or over-trust, and authority allocation, followed by future opportunities.


\textbf{{1. How to design the adaption law or adaptive/adaptable shared control? {Also, what} kind of role of the human driver should be in driver-vehicle shared control systems?}}

In Section III, we have discussed different ways to design adaptive shared control between the human driver and an automated driving agent. Most of them were from U-shape, noncooperative game theory, and the torque applied on the steering wheel by the human. The role of human drivers in highly automated vehicles can be defined as
\begin{itemize}
	\item Convertible role between leader and follower\cite{flad2017Cooperative} (game theory);
	\item Symbiotic relations with the automated driving agent (page 24 in \cite{lamnabhi2017systems});
	\item Being an active driver, passengers, or passive drivers, but they may still be required to take over control\cite{ohn2016looking};
	\item Or being parallel\cite{tan2019guidance,chen2018parallel}.
\end{itemize}
The different \textit{roles} of human drivers in driver-vehicle systems {result in} various shared-control paradigms. In terms of the methods researchers utilized, the shared control performance can be improved by considering individual characteristics, classifying the levels of human-automation interaction, and integrating with cognitive psychology.

{The U-shape control law only qualitatively describes the relations between drivers' workload and the needs for assistance as well as the driving performance, but not a quantitative expression. As a result, the adaption control laws derived from U-shape was greatly different, for example, in literature \cite{nguyen2017driver} and \cite{oufroukh2014integrated}. Many factors could cause the difference, such as the diversity in individuals' driving experience and physical/psychological status. } Classifying the type of human driver\cite{wang2017drivingstyle}, regarding their abilities and characteristics, and then designing a personalized driver model\cite{butakov2015personalized} {capable of describing and adapting} this driver's characteristics\cite{habib2017adaptation} could be an efficient way to tackle this kind of problems.

In terms of the game-theory-based adaptive law, researchers usually modeled human drivers by assuming that drivers had a perfect internal model\cite{cole2017occupant} for understanding and predicting vehicle states as the same with an automatic controller, that is, both agents had an identical, deterministic objective functions \cite{na2013linear,flad2017Cooperative,na2017application}. In the real world, however, humans driving is not always a deterministic process\cite{yang2010development}, but in nature, a stochastic and dynamic process\cite{wang2017learning,nechyba1998stochastic}, and even impaired behaviors (e.g., fatigue and drunk). Therefore, the stochastic driver behavior and divergences among drivers should be modeled and accounted in future work. Besides, the underlying relations (the \textit{role} of a human driver in the driver-vehicle system with shared control schemes) between human drivers and automated driving agents {remain} open-end.

In response to {the problems above}, classifying the types and levels of human-automation interaction \cite{parasuraman2000model,pacaux2011levels,endsley1999level} and increase the agent's adaptability \cite{pacaux2016adaptive} could be an applicable approach. The automation functions usually cover four types: information acquisition, information analysis, decision and action selection, and action implementation. Within each of these types, the automation is defined and treated as a continuous level from low (i.e., fully manual) to high (i.e., fully automatic). Human performance consequences in terms of types and levels constitute primary evaluative criteria for automation design. This approach has been used in driver assistance system design \cite{cai2012coordinating}.

Besides, Beetz, \textit{et al.} \cite{beetz2007cognitive} and Heide and Henning\cite{heide2006cognitive} proposed the idea of cognitive car separately --- a technological cognitive system that can perceive itself and its environment, as well as collect and structure information in an autonomous way. In the cognitive cars, some key issues remain regarding action implementation (i.e., function delegation): 
\begin{itemize}
	\item What kind of actions should be implemented?
	\item When to add the actions?
	\item How to implement the function delegation appropriately?
\end{itemize}
{To date, we are not able to answer the three questions systematically, but most research focuses on (1) the effects of haptic support systems on driver performance, (2) vehicle stability control for collision avoidance, and (3) active steering system with adaptive assisted systems. One of the potentially useful approaches} is to consider and built cognitive cooperation based on cognitive psychology, as mentioned in\cite{hoc2001towards}. Li, \textit{et al}. also holds that the cognitive cars will be a new frontier advanced driver assistance system for research\cite{li2012cognitive}. Besides, the Petri net modeling has been demonstrated as an alternative solution\cite{wu2011petri}.

\textbf{{2. What is the appropriate trust? {Alternatively,} would drivers take a seat as passive occupants, who fully trust their vehicles?}}

Over-reliance on automated driving and driver complacency are often problematic\cite{billings1996human}, thus resulting in, for example, a long reaction time\cite{eriksson2017takeover}. The driver, who fully and deeply trusts the ability of driving automation, therefore failed to intervene and take manual control even as the driver crashed the car\cite{lee1992trust,lee2004trust}. Fortunately, human drivers have an appropriate level of trust in the automated driving agent, such as being aware that  the automation system can better sense and faster response \cite{johns2016exploring} and displaying system situation awareness\cite{sonoda2017displaying}, but sometimes can also make mistakes. Driver performance can also be improved by effectively conveying the limitations of automated driving to human drivers \cite{rezvani2016towards,tijerina2016exploratory}. Besides, an appropriate, elaborate practice for drivers could mitigate the negative impact of over-trust in the automated driving agent on reaction time\cite{payre2016fully}. 

{Besides, when the desired trajectories derived from the driver and the automated driving agent are similar, the vehicle inputs from both of them locally differ but combine without conflict. However, when the desired trajectories of two agents are different, things will become intractable and bring a question: \textit{which one input should be trusted and exerted to the vehicle?} One potential way for tackling this issue is to develop a psychologically-grounded cognitive-physical model \cite{lamnabhi2017systems} capable of correctly describing and predicting the driver's desired trajectory from the operational level, tactical level, and strategic level by understanding information processes (neuroscience) and cognitive abilities (psychology) of the driver. The cybernetic driver models integrating visual (including anticipation and compensation) systems with neuromuscular systems (or motor processes) are an efficient way to benefit shared control.} The readers can refer to the review paper \cite{mars2017modelling} for details about the model structure and parameter identification. {Another potential way is to design a metric to evaluate and analyze the conflicts between a human driver and automated driving agent\cite{itoh2016hierarchical}. Many indicators have been developed to evaluate shared control performance, and they mainly concern four aspects: accuracy (e.g., path-tracking errors)\cite{mars2017modelling,sentouh2018driver,li2018shared}, safety (e.g., risk metrics such as time to lane crossing)\cite{li2019shared,li2018shared}, compatibility (e.g., existing conflicts or not)\cite{mars2017modelling,sentouh2018driver}, and robustness (e.g., the vibration in the resonance response of vehicle)\cite{sentouh2018driver,saleh2013shared}. In the haptic shared control system, the directions and periods of the torque applied to the steering wheel by a human driver and automated driving agent are most commonly used. For example, researchers in \cite{mars2017modelling} considered four aspects of steering torque to evaluate conflicts in haptic shared control: consistency rate, resistance rate, contradiction rate, and contradiction level.}

\begin{figure}[t]
	\centering
	\includegraphics[width = 0.48\textwidth]{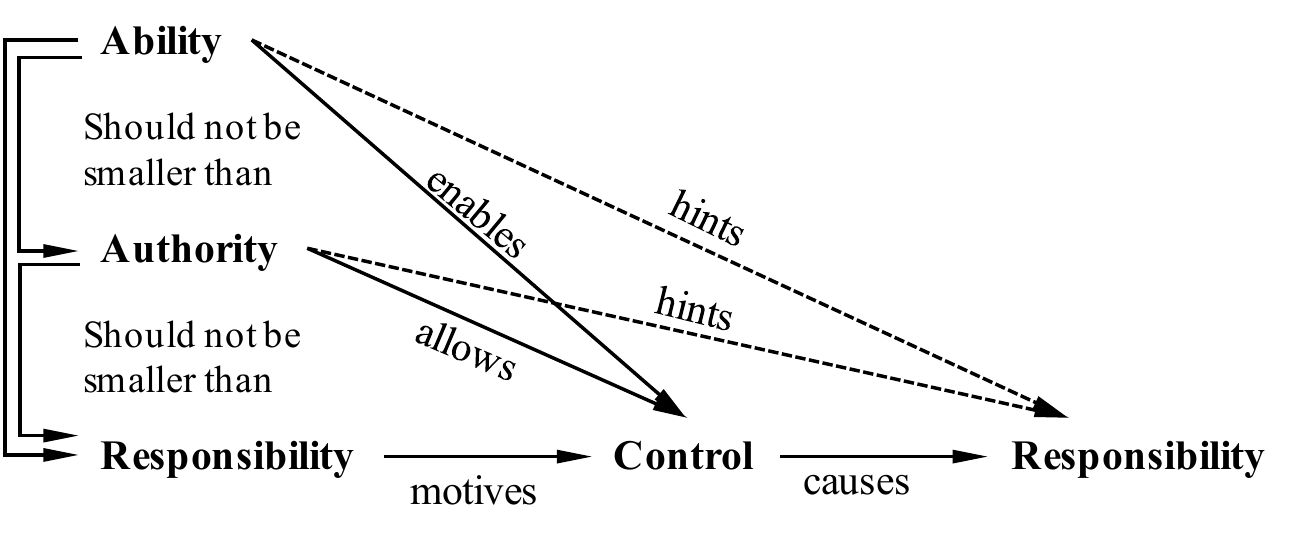}
	\caption{Relations between ability, authority, control and responsibility\cite{flemisch2012towards}.}
	\label{fig:relations}
\end{figure}

\textbf{{3. Which way is the best transition in authority, ability, responsibility, and control (A2RC)?}}

{Before presenting the open-ended questions}, we should define some basic concepts in interactions between human driver and automated driving agents according to \cite{flemisch2012towards}:
\begin{itemize}
	\item \textit{Authority}, which can be defined by what the human driver or automated driving agent is allowed to do or not to do. Further, the authority can be partially and continuously transferred between human drivers and automated agents.
	\item \textit{Ability}, which can be defined as the possession of the means or skill to perceive and select adequate action and act appropriately.
	\item \textit{Responsibility}, which can be assigned beforehand to motivate certain actions and evaluated afterward, where the human driver or automated driving agent is held accountable or to blame for a state or action of the driver-vehicle system and consequences resulting thereof.
	\item \textit{Control}, which means having the power to influence the vehicle states.
\end{itemize}
Fig. \ref{fig:relations} shows the relations between the four concepts. Based on these basic definitions, the issues that existed in human-automation interaction systems are also encountered in the human-vehicle shared control systems since the highly automated vehicle is a case-specific automation system\cite{flemisch2012towards,parasuraman2000model}, for example:
\begin{enumerate}
	\item How do we balance between exploiting increasingly powerful technologies and retaining authority for human drivers?
	\item How can we define clear, safe, efficient and enjoyable roles between human drivers and automated agents?
	\item Which of the subsystems of future human-vehicle systems should have which ability, which authority and which responsibility? Alternatively, which system functions should be automated and to what extent?
	\item What other concepts besides authority and responsibility do we need to describe and shape a dynamic but stable balance between human drivers and automated driving agent?
\end{enumerate}

According to these predefined terms, there is an allocation problem  in terms of A2RC between human drivers and automated driving agents. Most literature only focuses on the control authority between human drivers and automated driving agents, but ignore the relations regarding the A2RC. In {the} real world, the authority, ability, responsibility, and control are not independent. In \cite{flemisch2010towards}, an A2RC diagram was developed to analyze and design human-machine systems, which can be used to tackle the transition issues of A2RC in driver-vehicle shared control systems. Besides, Acarman, \textit{et al}. \cite{acarman2003control} also proposed a control authority transition system for collision and accident avoidance considering drivers' physical and cognitive capacities as well as the situation/danger/hazard analysis. In addition to the A2RC, the H-metaphor (H-mode or \textit{horse} metaphor)\cite{flemisch2003h}, {i.e., a} metaphor of the relationships between a horseback rider and the horse, could be as a  guide to open new horizons in the shared-control systems, which has shown its remarkable achievement in aircraft co-pilot design\cite{goodrich2006application} and highly automated driving \cite{flemisch2014towards}.

On the other hand, exploiting brain-related signals could directly offer rich information about human driver intent, ability and thus allow us to optimize the allocation of authority and reduce the conflict between two agents. A well-designed human-vehicle interaction interface integrated with human driver psychological and psychobiological characteristics \cite{card1983psychology,mars2016psychophysiology} and active capabilities\cite{flemisch2008cooperative} could benefit shared-control tasks by offering much cognition information. Moreover, the feedback from the interaction interface could also show if a poor-designed interface, adverse effects on human driver \cite{jaschinski2016impact} and control performance\cite{lobjois2016effects}. Similar to the human-computer interaction design (see pages 29-48 in \cite{mackenzie2012human}) from an empirical research perspective, human factors regarding perception sensors (i.e., vision, vestibular and kinesthetic, somatosensory, and auditory, as discussed in Section IV-\textit{A}-\textit{1)}), action responders (i.e., limbs and legs, as discussed in Section IV-\textit{A}-\textit{3)}), and brain factors (i.e., perception, cognition, and memory, as discussed in Section IV-\textit{A}-\textit{2)}) could be carefully considered and implemented based on the methods of human-robot shared control (see the review literature in \cite{goodrich2008human,lasota2017survey}). One of the typical applications is to use brain-related signals to enhance shared-control performance through the brain-vehicle interface \cite{lu2018mathematical,bi2017queuing,bi2014using,lu2019model}. Various brain-vehicle interfaces with different capabilities (e.g., adaptive brain-vehicle interface\cite{gandhi2014eeg}) have been designed based on EEG signals\cite{carlson2013brain}). More literature refers to the review literature \cite{bi2013eeg,abbink2018topology}.

Moreover, human factors (e.g., the way to shared information with the human driver) can influence the transitions of A2RC. An inefficient design of shared control interface would cause conflicts between human drivers and automated driving agent, even the catastrophic consequences. Eriksson \textit{et al}. \cite{eriksson2019rolling} investigated different ways to support driver decision-making during automation-to-manual transitions in a take-over scenario. Most related works have been comprehensively discussed in the recent review paper \cite{lu2016human}. However, it is still not entirely clear about the influence of different ways (such as torque, steering angle, and vibration) to information transition between a human driver and automated driving agents on the performance during shared control.

\section{Discussion and Conclusion}
The motivation for this review paper is to illustrate how to model a driver-vehicle shared control system and understand the challenges and opportunities for highly automated vehicles with human drivers still retained in the control loop. We have discussed the architectures of driver-vehicle shared control systems, the approaches to modeling the complex systems, and the future challenges and opportunities. We have provided a survey of the progress over the past decades in driver-vehicle shared control technologies. In order to understand the complex driver-vehicle systems, we decoupled it into different subsystems and summarized how to model them by reviewing the state-of-the-art literature. Finally, we have provided discussions on the challenges and opportunities in this field. While advanced driver assistance systems have been developed and introduced over the past decade, a deeper and more holistic understanding of the relationship between human drivers and automated driving agents and the way that human drivers cognitively interact with the driving environment will remain an active area of research in next few years.

\ifCLASSOPTIONcaptionsoff
  \newpage
\fi



\bibliographystyle{IEEEtran.bst}
\bibliography{bibtex/bib/MyBibReference}

\end{document}